# Anomalous compressibility of ferropericlase throughout the iron spin crossover

R. M. Wentzcovitch,[1,2,*] J. F. Justo,[1,2,4] Z. Wu,[1,2] C. R. S. da Silva,[2]
D. A. Yuen,[2,3]  and D. Kohlstedt[3]

[1]Department of Chemical Engineering and Materials Science,

[2]Minnesota Supercomputing Institute,

[3]Department of Geology and Geophysics,

University of Minnesota, Minneapolis, MN 55455

[4]Escola Politécnica, Universidade de São Paulo, CP 61548, CEP 05424-970,

São Paulo, SP, Brazil

The thermoelastic properties of ferropericlase $Mg_{1-x}Fe_xO$ ($x$ = 0.1875) throughout the iron high-to-low spin crossover have been investigated by first principles at Earth's lower mantle conditions. This crossover has important consequences for elasticity such as an anomalous bulk modulus ($K_S$) reduction. At room temperature the anomaly is somewhat sharp in pressure but broadens with increasing temperature. Along a typical geotherm it occurs across most of the lower mantle with a more significant $K_S$ reduction around 1400-1600 km depth. This anomaly might also cause a reduction in the effective activation energy for viscous creep and lead to a viscosity minimum in the mid-lower mantle, in apparent agreement with results from inversion of data related with mantle convection and postglacial rebound.

## Introduction

Understanding of the Earth's lower mantle relies on indirect lines of evidence. Comparison of elastic properties extracted from seismic models with computed or measured elastic properties of candidate minerals at mantle conditions is a fruitful line of enquiry. For instance, it has shed light on the lower mantle composition (1-3) and on the nature of the D" layer (4,5). Such comparisons support the notion that the lower mantle consists primarily of ferrosilicate perovskite, $Mg_{1-y}Fe_ySiO_3$, and ferropericlase, $Mg_{1-x}Fe_xO$ (hereafter Pv and Fp respectively). In contrast, evidences based on solar and chondritic abundances suggest a deep lower mantle chemical transition into a pure Pv composition at ~1000 km depth (6). A chemical transition with wide topography, gentle, and diffuse changes in elasticity and density is also supported by geodynamic modeling (7). The discovery of the spin crossover in Fp and in Pv at lower mantle pressures (8,9) introduces a new dramatic ingredient that demands a careful reexamination of these phases' elastic properties at appropriate conditions, the consequences for mantle elasticity, and reanalysis of lower mantle properties. This may, after all, support lower mantle models containing a chemical transition. Here we show the effect of the spin crossover on the bulk modulus and bulk velocity of Fp at high temperatures. We also show the effect it should have on the bulk modulus of a homogeneous lower mantle with pyrolite composition and confirm and justify the origin of anomalies in the elasticity of Fp recently demonstrated at room temperature (10). We point out that such elastic anomaly might alter the activation energy for viscous creep (11,12) in Fp which might affect mantle viscosity.

## Discussion and Results

The high spin (HS), S=2, to low spin (LS), S=0, crossover in ferrous iron in Fp has been detected by several techniques at room temperature (8, 10,13-17) and recently up to 2000 K (18). For typical mantle compositions the crossover may start as low as ~ 35 GPa (17) and end as high as 75 GPa (8) at room temperature. The observed variations in the pressure range of the transition seem to be related to the variable degree of hydrostaticity in experiments. This pressure range broadens substantially with increasing temperature (18). This is actually a crossover that occurs continuously (19,20) passing through a mixed spin (MS) state. Here, we extend the earlier thermodynamics formalism developed to investigate the spin crossover in Fp (20) by including the spin state dependent vibrational properties. Equations of state do not have predictive quality unless they include phonon related effects. This is a particularly challenging task given the strongly correlated nature of this solid solution. We then obtain the high temperature compressibility and bulk velocity of Fp. We also address the potential effect the anomalous compressibility across the spin transition might have on the creep viscosity of Fp.

**Thermodynamics of the crossover transition**

We treat Fp in the MS state as an ideal solid solution of HS and LS states. This approximation seems to be well justified by the concentration-independent static spin transition pressure for concentrations up to x=0.1875 (20). Therefore:

$$V(n) = nV_{LS}(P,T) + (1-n)V_{HS}(P,T) \qquad (1)$$

$$\frac{V(n)}{K(n)} = n\frac{V_{LS}}{K_{LS}} + (1-n)\frac{V_{HS}}{K_{HS}} - (V_{LS}-V_{HS})\frac{\partial n}{\partial P}\bigg|_{T} \qquad (2)$$

where $n=n(P,T)$ is the LS fraction, and $V_{LS}$, $V_{HS}$, $K_{LS}$, and $K_{HS}$ are the equilibrium volume and isothermal compressibility of pure LS and HS states. Eq. (2) differs from the weighted average of the compressibilities by an additional term caused by the pressure dependence of $n(P,T)$. This last term is ultimately responsible for the anomaly on the bulk modulus recently reported by Crowhurst et al. (10). According to Eqs. (1) and (2), the properties of Fp in the MS state may be determined from those of the LS and HS states plus the low spin fraction, $n(P,T)$, all of which must be computed by first principles.

In contrast to the previous thermodynamics treatment (20) we now include vibrational effects. It is impossible to address thermodynamics properties without them. The other approximations used in Ref. (20) to compute $n(P,T)$ are retained. They are:

1) The magnetic entropies are $S_{HS}^{mag}(n) = k_B X_{Fe} \ln[m(2S+1)]$ and $S_{LS}^{mag} = 0$ for the HS and LS states respectively.

2) The HS-LS configuration entropy is $S_{conf} = -k_B X_{Fe}[n \ln n + (1-n) \ln(1-n)]$. Fluctuations in $n(P,T)$ are insignificant given the finite sample sizes. Because in this solid solution configurations are not expected to be static, this formula implicitly assumes the ergodic hypothesis, i.e., time and ensemble averages are equal.

3) The Mg/Fe configuration entropy is insensitive to spin state.

$n(P,T)$ is then obtained by minimizing the Gibbs free energy with respect to $n$. This leads to:

$$n(P,T) = \frac{1}{1 + m(2S+1)\exp\left[\dfrac{\Delta G_{LS-HS}^{stat+vib}}{X_{Fe}K_B T}\right]} \tag{3}$$

where $\Delta G_{LS-HS}^{stat+vib}(P,T)$ is the difference between the static plus vibrational contributions to the free energy of the LS and HS states, $X_{Fe}$ is the concentration of iron (here 0.1875), $S$ and $m$ are respectively spin and orbital degeneracies of the HS ($S$=2 and $m$=3) and LS ($S$=0 and $m$=1) states. Therefore, to obtain $\Delta G_{LS-HS}^{stat+vib}(P,T)$ and $n(P,T)$ we need first to obtain the full vibrational spectrum and free energies of pure spin states within the QHA.

**The Vibrational Virtual Crystal Model (VVCM)**

The thermal properties of Fp in pure spin states were computed using the quasiharmonic approximation (QHA) (21) in which the Helmholtz free energy is given by:

$$F(V,T) = \left[U(V) + \sum_{qj} \frac{\hbar \omega_{qj}(V)}{2}\right] + k_B T \sum_{qj} \ln\left[1 - \exp\left(-\frac{\hbar \omega_{qj}(V)}{k_B T}\right)\right] \tag{4}$$

where $U(V)$ is the volume dependent static total internal energy obtained by first principles and $\omega_{qj}(V)$ is the corresponding volume dependent phonon spectrum.

Current methodological limitations preclude a direct computation of the vibrational density of states (VDOS) of Fp within the first principles LDA+U approach (see Methods). To circumvent this problem we developed a vibrational virtual crystal model (VVCM), within the same spirit of the virtual crystal (VC) model described in Ref. (22). The VC concept involves the replacement the atomic species forming the solid solution, in this case magnesium and HS or LS irons, by an "average cation" that can reproduce the properties of the solid solution. This approximation has been widely used in electronic structure calculations (23, 24). Here we develop a VC to compute only vibrational and thermodynamics properties. We are not aware of previous similar attempts in the literature. The development of successful VVCMs would be extremely useful to bypass the difficult problem of computing VDOS for numerous configurations involving hundreds of atoms representative of solid solutions, especially strongly correlated ones, so common in minerals.

The VVCMs corresponding to the pure HS and LS states consist of two atoms per cell in the rocksalt structure: oxygen and a virtual (cation) atomic species with a mass

$$M_{VC}^{cation} = (1 - X_{Fe}) M_{Mg} + X_{Fe} M_{Fe} \qquad (5)$$

where $M_{Mg}$ and $M_{Fe}$ are respectively the atomic masses of magnesium and iron, with the latter being independent of the iron's spin state. The interactions of the VC cation in the solid are modeled for the purpose of computing vibrational and thermodynamics properties only.

The VVCMs are essentially periclase, MgO with modified interatomic force constants that reproduce the elastic constants of HS or LS Fp and cation masses as in Eq. (5). The force constants of periclase were previously computed by first principles and produce excellent phonon dispersions (25,26). The force constants of the HS or LS VCs are obtained by matching the elastic constants extracted from the acoustic phonon dispersions (27-30) close to k = 0 to the elastic constants of HS and LS calculated by first principles. There is a linear relationship between force constants $D_{\mu\nu}(R^{ij})$ and elastic constants $C_{\sigma\tau\alpha\beta}$ (31):

$$C_{\sigma\tau\alpha\beta}(V) = \sum_{(i,j),(\mu,\nu)} a_{\mu\nu,\sigma\tau\alpha\beta}^{ij}(V) D_{\mu\nu}^{ij}(V) \qquad (6)$$

Here, Greek letters refer to Cartesian indices, $C_{\sigma\tau\alpha\beta}(V)$ are the volume dependent elastic constants in cartesian notation, while $D_{\mu\nu}^{ij}$ are the interatomic force constants between atoms $i$ and $j$ separated by $\boldsymbol{R}^{ij}$ when displaced in directions $\mu$ and $\nu$ respectively. The sum in Eq. (6) is over all atomic pairs $(i,j)$ and $a_{\mu\nu,\sigma\tau\alpha\beta}^{ij}(V)$ are a set of volume dependent constants. Due to symmetry constrains, many of the $a_{\mu\nu,\sigma\tau\alpha\beta}^{ij}$ constants vanish. Eq. (6) is a convergent summation since the force constants vanish rapidly with the interatomic distances. The convergence in Eq. (6) is guaranteed if the force constants vanish faster than $1/R^5$, where $R$ is the interatomic separation (31).

The force constants defined as

$$D_{\mu\nu}^{ij} = \frac{\partial^2 E(R^{ij})}{\partial R_i^\mu \partial R_j^\nu} \tag{7}$$

are used to compute the phonon spectrum at each volume:

$$\det \left| \frac{D_{\mu\nu}^{ij}(V)}{\sqrt{M_i M_j}} - \omega^2 \right| = 0 \tag{8}$$

and we need to obtain $D_{\mu\nu}^{ij}(V)$ for HS and LS VVCMs. Fp and periclase have only three elastic constants, $C_{11}$, $C_{12}$ and $C_{44}$ (Voigt notation) (32). We may modify three force constants of periclase independently to reproduce the static elastic constants of HS and LS Fp. We modified the three largest interatomic force constants of periclase, $D_{xx}^{12}$ (Mg-O nearest neighbor longitudinal interaction), $D_{xy}^{11}$ (Mg-Mg nearest magnesium interaction), and $D_{xy}^{12}$ (the Mg-O nearest neighbor transverse interaction). All other force constants of MgO are least one to two orders of magnitude smaller and remained unchanged. Changes in those force constants have only a minor effect on the elastic constants. More details of the VVCM developed here will be reported somewhere else (29,30).

A comparison between the static bulk modulus obtained by fitting an equation of state to the energy versus volume relation in HS and LS $Mg_{1-x}Fe_xO$ ($x$=0.1875) and the bulk modulus obtained from the elastic constants of the respective VCs is shown in SFig. 1. The virtual crystals produce distinct vibrational density of states (VDOS) for periclase, HS, and LS Fp (see SFig. 2). The acoustic mode dispersions of the HS and LS VVCMs are precisely the same as those of HS and LS Fp. This insures that thermodynamics calculations are carried out with the correct VDOS at low frequencies, which matter the most, and with a reasonably good weight averaged VDOS at high frequencies as well. The VVCM should offer more accurate thermodynamics properties than a Debye-like model because of the more detailed structure of the VDOS. We considered carefully the pressure/temperature range of validity of the QHA. Full and dashed lines in all figures correspond to conditions within and outside its range of validity, respectively. The upper temperature limit of the QHA is adopted as the lowest temperature of the inflection points in the thermal expansivity of pure LS and HS Fp at every pressure (1,33), i.e., $\partial^2 \alpha / \partial T^2|_P$=0.

The low spin fractions, $n(P,T)$, obtained from Eq. 3 is shown in Fig. 1. The black and white lines correspond to $n(P,T) = 0.5$ in calculations that include and exclude, respectively, the vibrational contribution to the free energy. Inclusion of the vibrational energy shifts the center of the crossover pressure range to higher pressures at higher temperatures and improves the agreement with experiments (18). The non-monotonic behavior of the measured LS population equal to 50 % (18) still needs to be verified by further experiments and the discrepancy between predictions and measurements should not be taken too seriously at this point. At lower temperatures our results agree best with results by Fei et al. (17) and underestimate the transition pressure compared to other experiments. The degree of hydrostaticity may vary in these experiments but we don't exclude the possibility we might be underestimating the transition pressure. We are probably underestimating iron-iron interactions because of the uniform iron distribution in the supercell adopted in the calculations. It is known that the crossover pressure increases with increasing $X_{Fe}$ due to iron-iron interaction (17, 34). Therefore, calculations with randomly distributed irons in Fp might produce broader pressure ranges at higher pressures, even at 0 K, even though iron cations avoid each other owing to elastic-type interactions caused by strong Jahn-Teller distortions (20). After inclusion of zero point motion effects, our new transition pressure at 0 K is 36 ± 3 GPa. Because the amplitudes of the anomalies in $K_S$ and $V_{\phi}$ increase with decreasing crossover pressure range, our results should be an upper bound of these values. However, our predicted anomalies should be more accurate at mantle temperatures where the agreement with experimental data improves (18) (see Fig. 1). The latter might be a signal of improved hydrostaticity in the high temperature data.

The isothermal compression curves throughout the MS state are shown in Fig. 2. The 300 K isotherm displays an anomaly that is somewhat more pronounced than the experimental ones (14,17). Our transition region agrees better with Fei *et al.*'s data (17) on Fp with $X_{Fe} = 0.2$. Two data sets were published: the data up to ~64 GPa agrees better with our predictions for the HS state while data up to ~95 GPa agrees better with the LS state. The difference with respect to Lin et al.'s data (14) correlates with the difference between our predictions and their measurements (18) of the LS fractions at low temperatures (see Fig. 1). The theoretical volume at 300K and 0 GPa, 11.46 cm$^3$/mol, is slightly larger than the experimental one, 11.35 cm$^3$/mol, on samples with $X_{Fe}$=0.17 (14,18) and slightly smaller than that on samples with $X_{Fe}$=0.2 (17), 11.51 cm$^3$/mol. This is consistent with the difference in concentrations ($X_{Fe}$=0.1875 here). The theoretical volume reduction due to the spin collapse is in average 4% for the entire pressure range, compared to about 2-4% by experiments with similar compositions (14-18). The experimental extrapolation of the HS equation of state to the LS stability field may contribute to some uncertainty in the estimated experimental volume reduction. At high temperatures the anomaly becomes almost imperceptible and may be difficult to detect experimentally at lower mantle temperatures and pressures. Lin and Tsuchiya (35) also reported calculated compression curves at high temperatures. Besides the strategy, the main difference between those calculations and ours is the choice of the LS fraction, *n(P,T)*, and the treatment of vibrational effects. They they adopted the same VDOS for MgO, HS and LS Fp. Besides, they used *n(P,T)* reported in Ref. (20) which is thermodynamically inconsistent once the phonon contribution to the free energy is included. The major difference between those results and ours is essentially captured in Fig. 1. Inclusion of composition and spin dependent VDOS shifts the crossover pressure range to higher pressures at higher temperatures and broadens the transition pressure range (compare white and black lines in Fig. 1). Therefore the anomalies in Ref. (35) are more enhanced and occur at lower pressures.

The quality of our predictions can also be tested by inspecting the thermal expansivity, $\alpha$, shown in Fig. 3. At low or very high pressures $\alpha$ has normal behavior since the system remains respectively in pure HS or LS states. The normal thermal expansivity of the HS state is essentially the same as that of MgO (36,37) as observed experimentally (37). Within the range of validity of the QHA, the magnitude of $\alpha$ at 0 GPa also agrees very well with measurements for other concentrations (36). Throughout the spin crossover $\alpha$ behaves anomalously also. This type of anomaly has been measured in LaCoO$_3$ perovskite, another system that undergoes a spin crossover transition (perhaps more than one) with increasing temperature at 0 GPa (38). The amplitude of the anomaly may be somewhat overestimated in our case because of the perhaps narrower calculated crossover range. Nevertheless, this effect could have significant consequences for the mantle geotherm, mantle dynamics, and for temperature induced lateral heterogeneities if the spin transition indeed occurs in the mantle. Anomalies on several other thermodynamics quantities will be reported elsewhere (29).

The adiabatic bulk modulus, $K_S$, density, $\rho$, and bulk velocity, $V_{\Phi}$, along several isotherms are shown in Fig. 4. Below 35 GPa, our calculated $K_S$ and $\rho$ are in excellent agreement with the isothermal bulk modulus, $K_T$, and $\rho$ measured at 300 K in the HS state for $X_{Fe} = 0.17$ (14). Inclusion of vibrational effects improves considerably the agreement with experiments. The remaining difference is consistent with the difference in iron concentration. There is a considerable reduction in $K_S$ and $V_{\Phi}$ throughout the spin crossover that is consistent with the reduction in bulk modulus of Fp with $X_{Fe} = 0.06$ (10) shown in the same figure. The difference in the magnitude of the anomaly is also consistent with the difference in iron concentration, i.e., approximately a factor of 3. The magnitude of the anomaly is more noticeable at low temperatures: at 300 K the crossover pressure range is ~36-48 GPa compared to the experimental one, ~ 35-50 GPa (17), or ~50-75 GPa (8,10,14-16,18). Therefore, the difference between predictions and experimental data is comparable to differences between experiments, but our results agree particularly well with Fei *et al.*'s data (17).

**Potential effect of the spin crossover transition in Fp on the mantle bulk modulus**

The effect of the spin crossover in Fp along a typical geotherm (39) is shown in Fig. 5a,b. The anomalies in $K_S$ and $V_\phi$ start at ~40 GPa (~1000 km depth) and are most pronounced around 70 ± 20 GPa (1600 ± 400 km). However, the crossover continues down to the core-mantle boundary (CMB) pressure with a possible reentrance into the HS state due to the thermal boundary layer above the CMB (39). In contrast, density increases smoothly throughout the entire pressure range of the lower mantle. The shaded areas correspond to possible values of these quantities due to uncertainties in the calculated static transition pressure and the narrower range of our transition pressure.

The net effect of the spin transition in Fp on the bulk modulus of a uniform aggregate with pyrolite composition (40) along a mantle geotherm (39) is shown of Fig. 5c. This comparison is made to elucidate and highlight an effect that may be quite subtle. We have adopted the first principles bulk modulus of Pv reported earlier by our group (1). The effect of iron on the bulk modulus of Pv without the effect of its own spin crossover was included as reported in Ref. (28), $K(x) = K_0(1 + bX_{Fe})$, where $b$ varies linearly between 0.079 and 0.044 from 0 GPa to 136 GPa, respectively. Experimentally, Pv's bulk modulus is not noticeably affected by the spin crossover (41). Theory predicts LS ferrous iron to be displaced from the equilibrium HS site, and the volume change to be quite insignificant throughout the crossover (42). At 0 GPa the aggregate consists of 80 w% of $Mg_{(1-x)}Fe_xSiO_3$, with x=0.12, and 20 w% of $Mg_{(1-y)}Fe_yO$, with y=0.1875. This translates into a monotonic increase in V% of Pv in the lower mantle, from 79.6 V% to 80.8 V% from 23 GPa to 120 GPa. The bulk modulus of the aggregate was computed using the Voigt-Reuss-Hill average (43). Compared to PREM's bulk modulus ($K_{PREM}$) (44), the aggregate shows a subtle undulation which appears to be smoothed or cut through by PREM. The effect of the spin crossovers in Pv still needs to be better understood and more sensitive strategies need to be devised to identify the signature of this crossover in Fp which is a subtle one at lower mantle conditions. Nevertheless, given the accuracy of current predictions and the fact that PREM is a one dimensional model, $K_{PREM}$ does not appear to be inconsistent with $K_S$ of a uniform pyrolite aggregate with a spin crossover in Fp along a typical adiabatic geotherm. The effect may be more visible on along slabs with colder geotherms and even less noticeable along superadiabatic hotter geotherms.

## Correlation between mantle viscosity structure and the spin crossover in Fp

The softening of $K_S$ in Fp might also impact on mantle viscosity. Combination of a thermal convection model using Newtonian viscous flow and seismic tomography data have implied the existence of a local minimum in mantle viscosity centered around 1500 km (45,46). We notice the proximity of the viscosity minimum and of the predicted anomaly in the bulk modulus of Fp in the mantle (Fig. 5a). As a relatively minor, weaker phase comprising ≤20 V% of Earth's lower mantle, the influence of Fp on viscosity depends critically on its distribution. In a poorly-mixed system, Fp grains will be isolated from one another by Pv grains, which have a viscosity $\sim 10^3$ times that of Fp far from the spin crossover (47). With Pv forming a load-bearing framework, the effect of Fp on viscosity will be modest. However, if phase separation occurs during large-strain deformation, Fp will markedly impact lower mantle viscosity. Recent shear deformation experiments on partially molten rocks, as well as on two-phase rocks in which the viscosity of the two phases are significantly different, demonstrate a profound segregation of the constituent phases (48,49). Mineralogical segregation and compositional layering are also observed in highly strained, naturally deformed rocks (50). Bands rich in Fp, separated by regions rich in Pv, are thus anticipated in a deforming lower mantle. Once phase separation occurs, strain localizes in the weak, Fp-rich layers causing a significant decrease in viscosity relative to the viscosity of a homogenously mixed, two-phase rock (51).

Here we invoke an elastic strain energy model (ESEM) (11) for viscosity to estimate the potential impact of the bulk modulus softening on Fp's viscosity, $\eta_{Fp}$. A Newtonian sub-solidus flow is assumed consistent with a diffusion creep deformation mode expected in the mantle and with the model used to infer lower mantle viscosity on the basis of convection related and postglacial rebound data (45,46). Fp's viscosity, $\eta_{Fp}$, is then:

$$\eta_{Fp} = A \exp\left(\frac{G^*}{k_B T}\right) \qquad (9)$$

where $G^*$ is the extrinsic activation energy for the dominant deformation mechanism, i.e. ionic diffusion, and $A$ is a constant. At a depth $z$, $\eta_{Fp}(z)$, should be (12):

$$\eta_{Fp}(z) = \eta_{Fp}(z_0) \exp\left(\frac{G^*(z)}{k_B T(z)} - \frac{G^*(z_0)}{k_B T(z_0)}\right) \qquad (10)$$

where $\eta_{Fp}(z_o)$ and $G^*(z_0)$ are Fp's viscosity and activation energy at a reference depth $z_o$, here assumed to be the top of the lower mantle.

The ESEM relates the activation energy for diffusion, $G^*(z)$, with the shear and bulk modulus of the system. The ionic diffusion process induces bond stretching and/or shearing depending on the diffusion path. As such, the diffusion barrier is related to different extents to shear and bulk modulus. This is usually described as a parameterized dependence on the pure shear and dilatational contributions, $G_s^*(z)$ and $G_D^*(z)$, to the activation energy,

$$G^*(z) = \delta G_s^*(z) + (1-\delta) G_D^*(z) \qquad (11)$$

where $\delta$ is a free parameter. The other quantities are (12):

$$\frac{G_s^*(z)}{G_s^*(z_o)} = \frac{V(z)\mu(z)}{V(z_o)\mu(z_o)} \qquad \text{and} \qquad \frac{G_D^*(z)}{G_D^*(z_o)} = \frac{V(z)K(z)}{V(z_o)K(z_o)} \qquad (12)$$

with $\mu(z)$, $K(z)$, and $V(z)$ being shear and bulk moduli and volume, respectively. This model works well for metals but the relationship between the diffusion barrier and the elastic moduli for ionic systems may not be this simple, even though there are indications that this model describes well the high pressure and high temperature behavior of diffusion in MgO (54). Nevertheless, this model expresses a relationship that is very likely to exist in some similar form between viscosity and elastic moduli. Despite consistent experimental (54) and first principles results of migration barriers in MgO (55-57), similar investigations in Fp are still necessary to clarify this point. Much less is known about the shear modulus at this point. Room temperature measurements (10) have indicated that the shear modulus also softens throughout the spin crossover, but this has not been confirmed by theory (30) or by more recent Brillouin scattering data (58). Therefore, the situation remains controversial and shear deformation may enhance or damp the bulk modulus related viscosity anomaly. Experimental data (54) and modeling results have suggested $G^*(z_0) \sim$ 300-330 kJ/mol at uppermost lower mantle conditions ($z_o$ = 660 km, P = 23 GPa, T = 2000 K). We then assume $G^*(z_0)$ = 315 kJ/mol (54).

The impact of the softening of $K$ on Fp's viscosity, predicted by a purely dilatational ESEM ($\delta$=0 in Eq. (11)) is shown in Fig. 6 compared with the relative changes in lower mantle viscosity, $\eta(z)$ (45,46), with depth. All profiles have accentuated minima at ~1400-1600 km. The decrease in Fp's viscosity near the CMB in our model is caused by the reentrance into the HS state owing to the thermal boundary layer (39) above the CMB, while the more drastic reduction in mantle viscosity beyond 2000 km may be related with numerous additional factors (59), such as the approaching post-perovskite transition, the temperature profile. It appears to depend also on the inversion model used to obtain the viscosity (45,46).

This bulk modulus anomaly in Fp may not only affect the viscosity and dynamics of the mantle (60) but also its overall state (61) and properties. In general, it is anticipated that properties of Fp related with ionic diffusion, such as ionic conductivity, should improve in the MS state owing to its enhanced compressibility (anomalously "soft" bonds), even though ionic conductivity is not the prevailing electrical conduction mechanism at conditions explored so far (62-64). In contrast, such properties should deteriorate in the LS state compared to the HS state because of the reduction in lattice parameter. Heat (lattice) conductivity, instead, is expected to follow the opposite trend: it should be boosted in the LS state and damped in the MS state in comparison with the HS state. The spin crossover in Fp will need to be investigated from several different angles before a clearer picture of its consequences emerges.

## Methods

The static first principles results for the pure HS and LS states reported here are very similar to those reported in Ref. (20). The electron-ion interactions were described by ultra-soft pseudopotentials. The oxygen pseudopotential was generated by the Troullier-Martins method (65), in a $2s^2 2p^4$ configuration with local-$p$ orbital, and core radii $r(2s) = r(2p) = 1.45$ *a.u.* The magnesium pseudopotential was generated using the von Barth-Car method (66), with five different electronic configurations ($3s^2 3p^0$, $3s^1 3p^1$, $3s^1 3p^{0.5} 3d^{0.5}$, $3s^1 3p^{0.5}$, and $3s^1 3d^1$, respectively with weights of 1.5, 0.6, 0.3, 0.3, 0.3, 0.2) with local-$d$ orbital, and core radii $r(3s) = r(3p) = r(3d) = 2.5$ *a.u.* The iron pseudopotential was generated using the modified RRKJ method (67), in a $3d^7 4s^1$ configuration, with core radii of $r(4s) = (2.0, 2.2)$, $r(4p) = (2.2, 2.3)$, and $r(3d) = (1.6, 2.2)$ *a.u.*, where the first value represents the norm-conserving core radius and the second one the ultrasoft radius. The electronic wavefunctions were expanded in a plane wave basis set, where a cutoff of 70 *Ry* provided converged results. The Brillouin zone was sampled by a 2×2×2 grid of $k$-points.

The calculations used a rotationally invariant version of the local density approximation adding a Hubbard potential (LDA+U), where U is computed by an internally consistent procedure (68). The values of U used here are the same as those used in Ref. (20), where the dependence of U with supercell size, spin state, and pressure were carefully investigated. Atomic positions were always fully relaxed with forces determined by the LDA+U energy functional.

Calculations were performed in a supercell with 64 atoms for the concentration $X_{Fe} = 0.1875$ (24 Mg, 32 O, and 6 Fe), with substitutional ferrous iron in the magnesium site. Irons were positioned in a way that maximized the inter-iron distances within the supercell. With such iron distribution, $X_{Fe} = 0.1875$ is the upper concentration limit for which iron-iron interactions are negligible in the calculation.

Vibrational frequencies in a 4x4x4 $q$-point grid of the rocksalt structure were computed by linear response theory (69). Force constants were then obtained and used to compute dynamical matrices on a 16x16x16 $q$-point grid from which the VDOS was obtained.

*Acknowledgements:* We thank Kei Hirose for helpful comments. RMW thanks JSPS for support and the Department of Earth and Planetary Sciences of TIT for hospitality during the final stage of preparation on this manuscript. Calculations were performed with the Quantum ESPRESSO package (70) at the Minnesota Supercomputing Institute and on the Big Red Cluster at Indiana University. This research was supported by grants NSF/EAR 0635990, NSF/ATM 0428774 (***VLab***), NSF/DMR 0325218 (ITAMIT), and UMN-MRSEC.

# Figure captions

**Figure 1**. (P,T) diagram of the fraction of LS irons (*n*) along the spin crossover in Fp with $x = 0.1875$. Black and white lines correspond to $n(P,T) = 0.5$ including or not the vibrational free energy, respectively. The plus symbols are experimental data for $n = 0.5$ from Ref. (18) for $x_{exp} = 0.17$.

**Figure 2**. Isothermal compression curves of $Mg_{1-x}Fe_xO$ ($x = 0.1875$). Full (dashed) lines correspond to results within (outside) the (P,T) regime of validity of the QHA. + are experimental results for $x_{exp} = 0.17$ (18), ○ and Δ are two different runs to 64 GPa and 95 GPa respectively for $x_{exp} = 0.2$ (17).

**Figure 3**. Thermal expansivity of Fp along several isobars. Full (dashed) lines correspond to results within (outside) the (P,T) regime of validity of the QHA (1,33). Circles and crosses are experimental values at 0 GPa for $Mg_{1-x}Fe_xO$ with $x=0.0$ (36) and $x=0.36$ (37) respectively.

**Figure 4**. Pressure dependence of the calculated (a) adiabatic bulk modulus, $K_S$, and (b) bulk wave velocity, $V_\phi$ and density, $\rho$, of $Mg_{1-x}Fe_xO$ ($x = 0.1875$) along several isotherms. Full (dashed) lines correspond to results within (outside) the (P,T) regime of validity of the QHA (1,33). Experimetal data for $K_S$ on a sample with $X_{Fe} = 0.06$ (10) is shown in 7a. The calculate anomaly is approximately three times larger than the observed one. Crosses on 7b are experimental data at 300 K on a sample with $X_{Fe} = 0.17$ (14).

**Figure5**. Properties of $Mg_{1-x}Fe_xO$ ($x = 0.1875$) along a lower mantle geotherm (39): (a) adiabatic bulk modulus, $K_S$, (b) bulk velocity, $V_\phi$, and density, $\rho$, (c) and bulk modulus of an aggregate with pyrolite composition, $K_{pyr}$, compared with PREM's (44) bulk modulus, $K_{PREM}$. In 4a and 4b, dashed and dotted lines correspond to properties computed in the MS, HS, and LS states respectively. Grey shaded regions represent the uncertainties derived mainly from the uncertainty in the computed enthalpies of HS and LS states and spin crossover pressure at T = 0 K.

**Figure 6**. Viscosity of ferropericlase along a mantle geotherm (39) compared with mantle viscosity models (46,47). Thin and thick black lines (46) were derived from convection related data extracted from geoid inversion and two distinct tomography models, Ref. (52) and (53) respectively. Thin and thick light brown lines (47) included also glacial rebound data in the derivation of the viscosity models. The tomography models were the same as those used in Ref. (46). Thick maroon line is the relevant property of Fp computed in the MS state. $\beta_D(z)$ is the contribution of the dilatational component of the activation energy, $G_D(z)$ , to $\eta_{Fp}(z)$ ($\delta$=0 only).

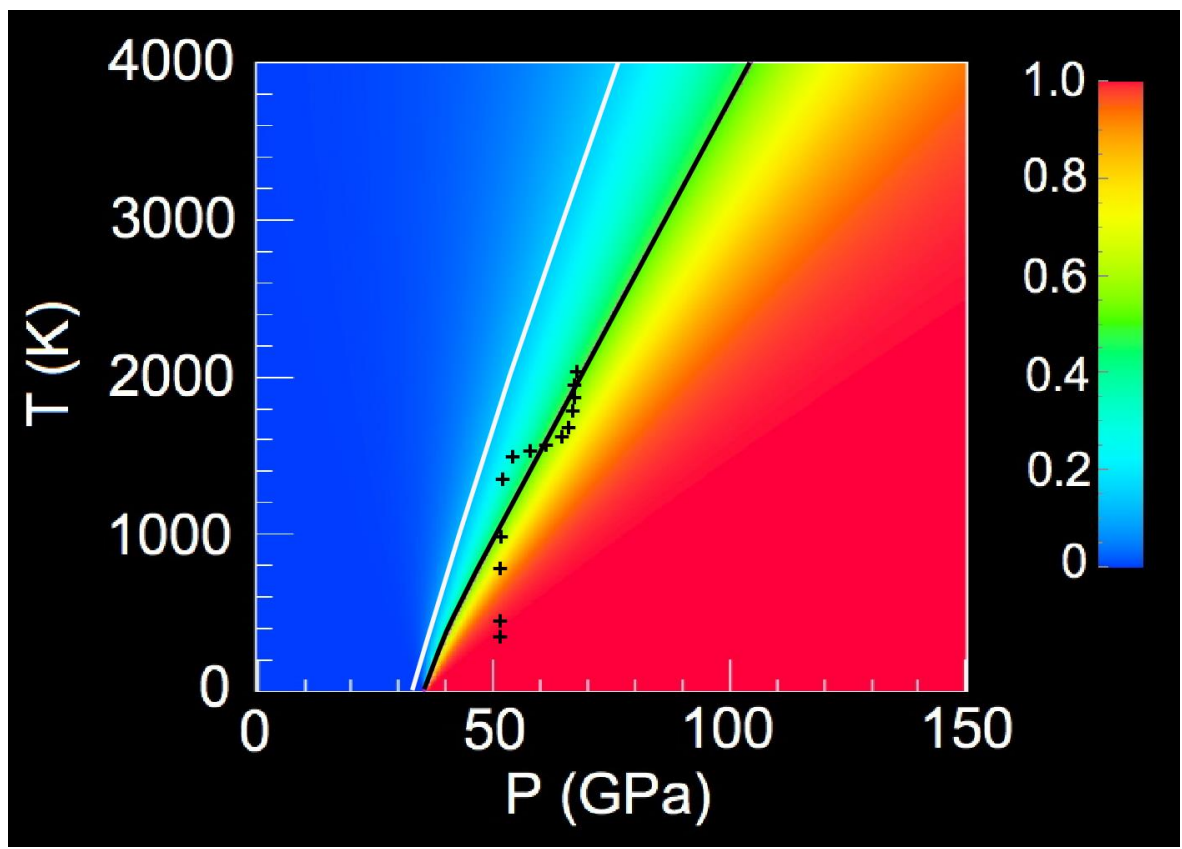

Figure 1

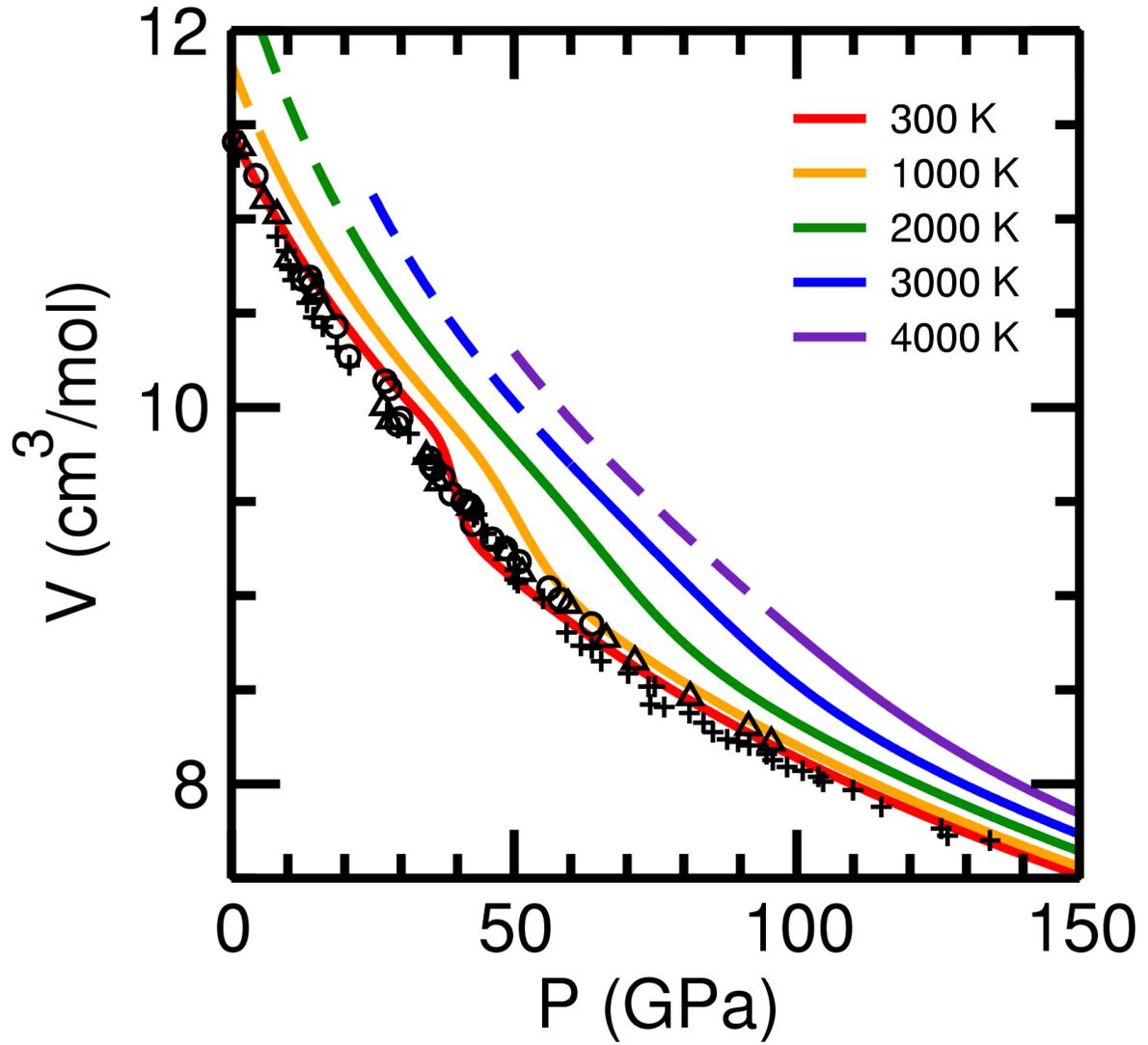

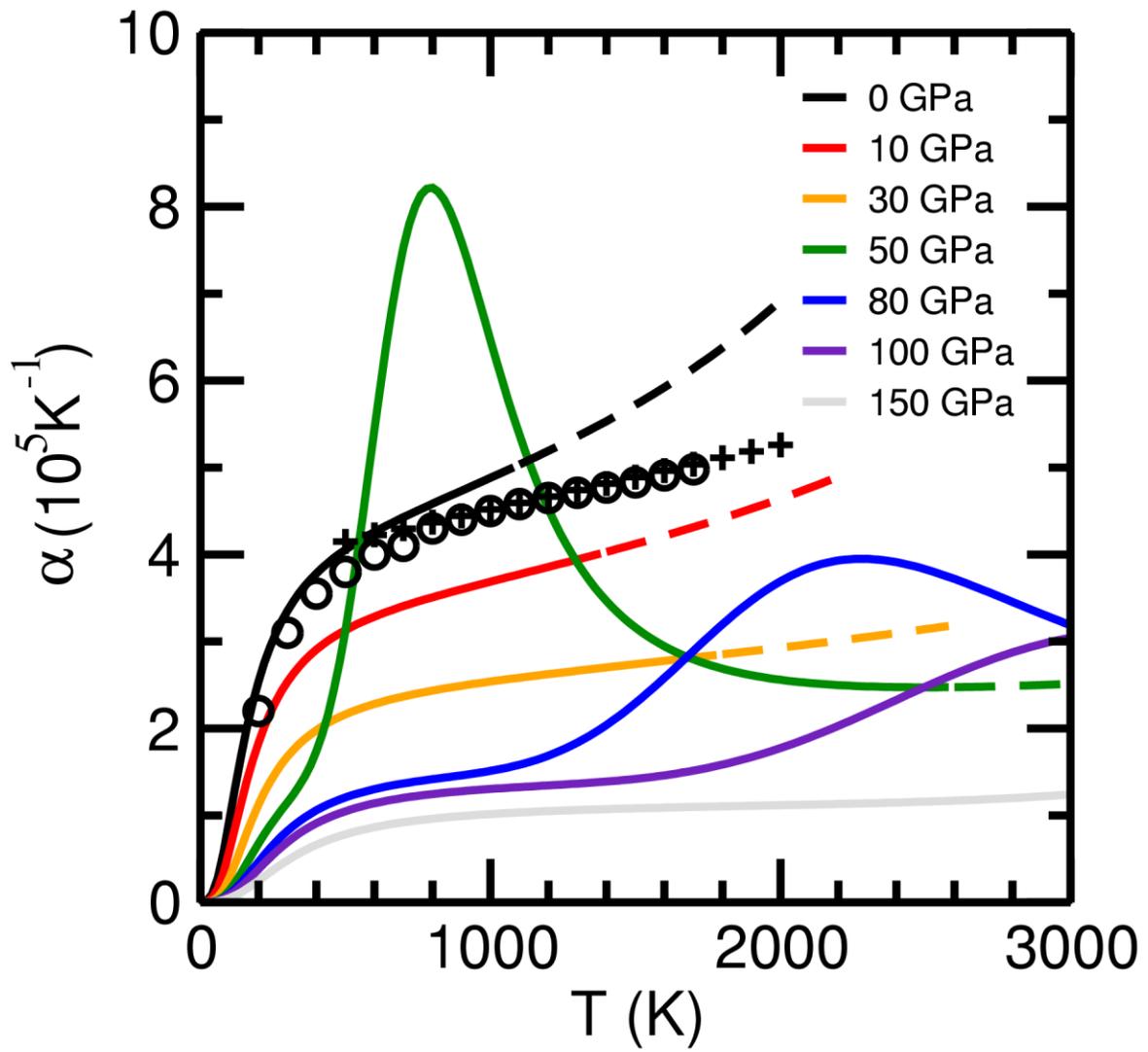

Figure 3

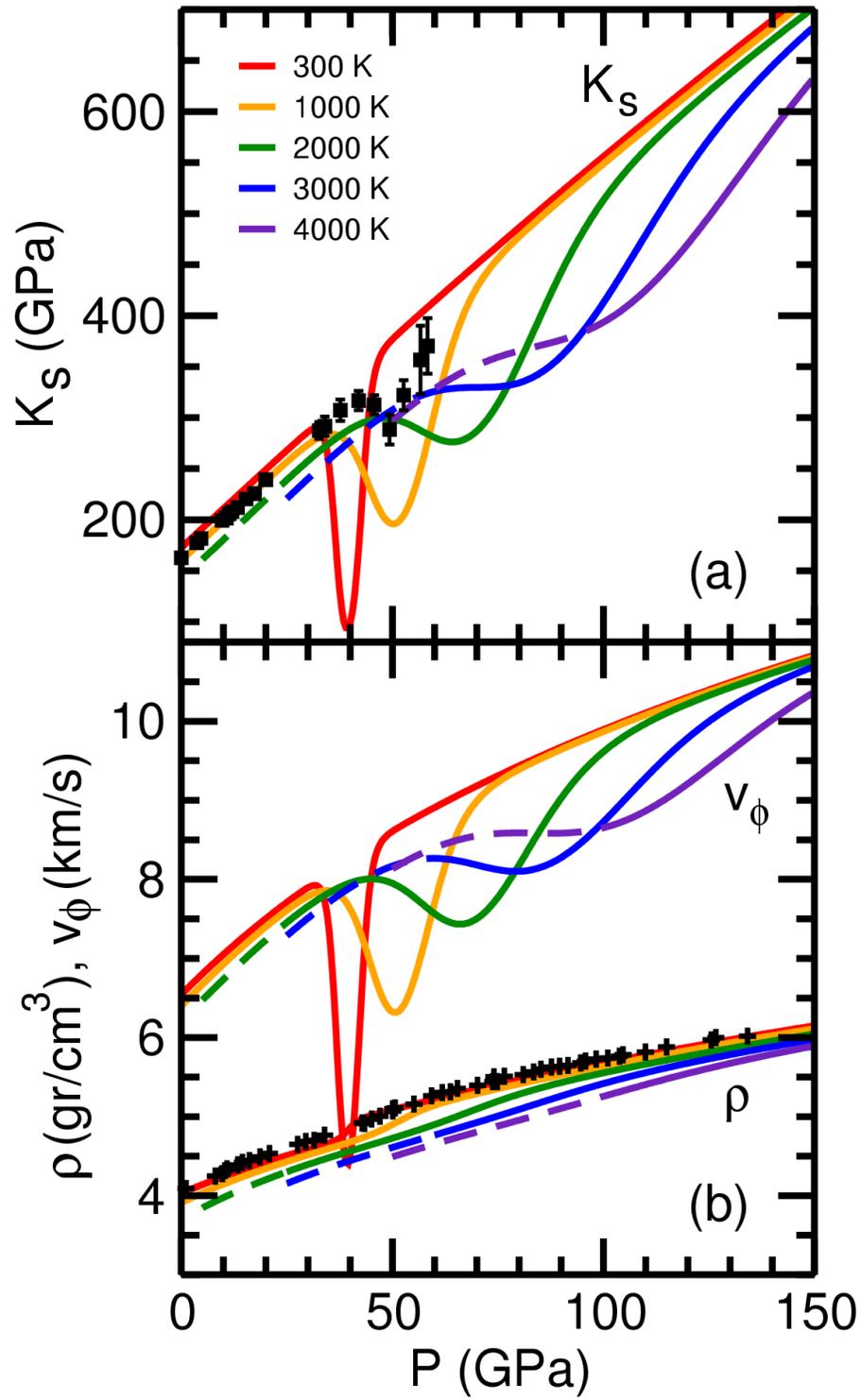

Figure 4

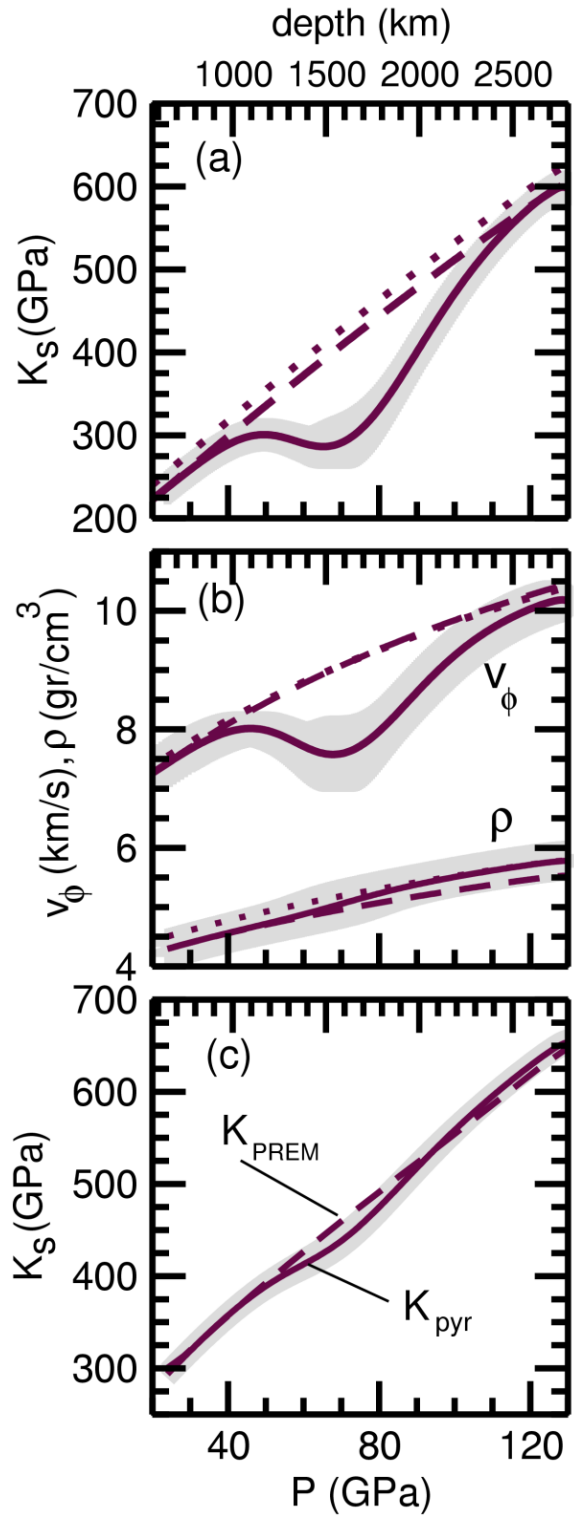

Figure 5

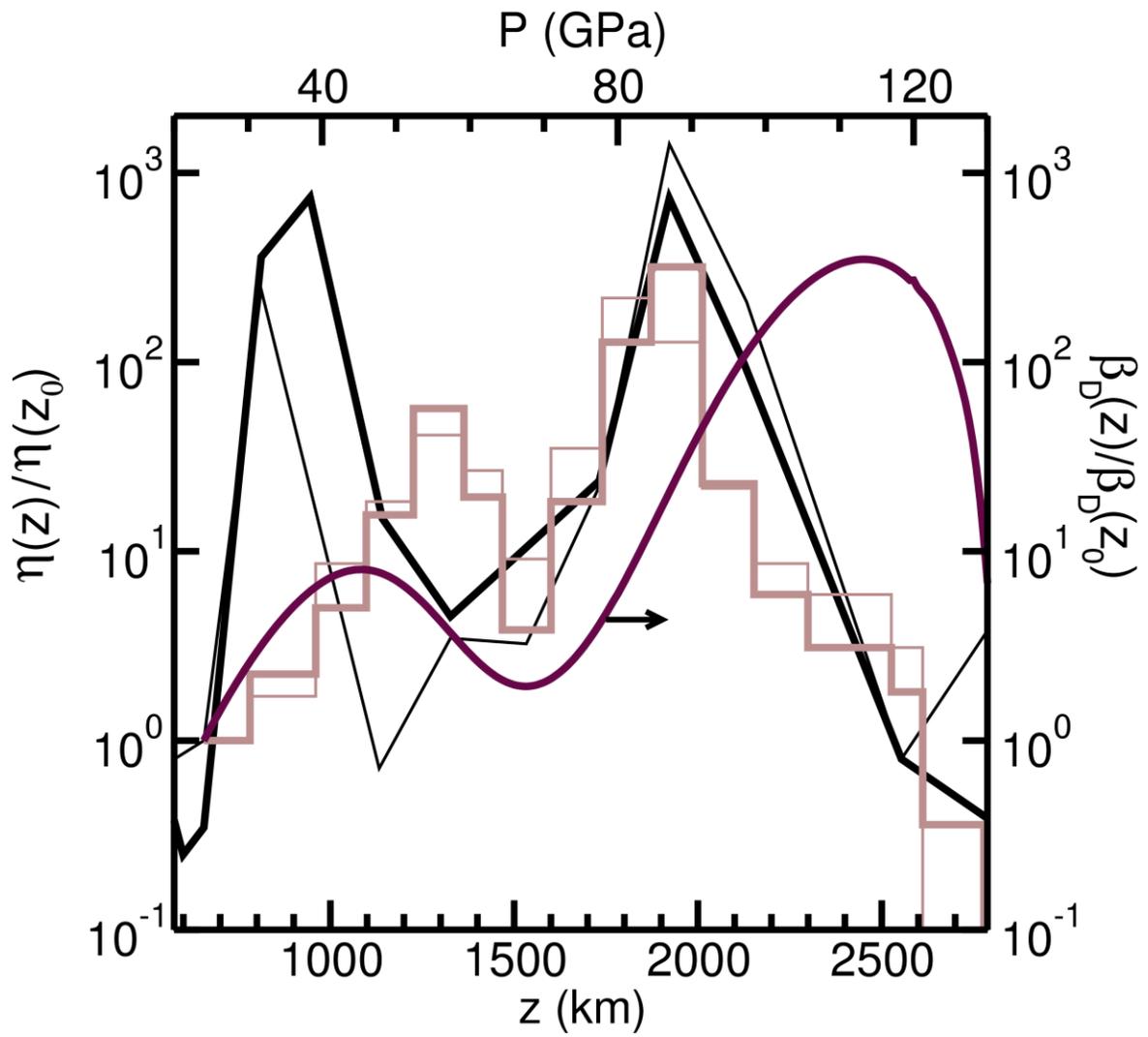

Figure 6